\documentclass[aps,prb,twocolumn,showpacs,superscriptaddress]{revtex4-1}

\usepackage{natbib}
\usepackage{amssymb}
\usepackage{amsmath}
\usepackage{graphicx}
\usepackage{dcolumn}
\usepackage{bm}
\usepackage{color}

\begin{document}

\title{Impurity induced enhancement of perpendicular magnetic anisotropy in Fe/MgO tunnel junctions}
\date{\today}
\author{A.~Hallal}

\author{B.~Dieny}

\author{M.~Chshiev}
\affiliation{Univ. Grenoble Alpes, INAC-SPINTEC, F-38000 Grenoble, France}
\affiliation{CNRS, INAC-SPINTEC, F-38000 Grenoble, France} 
\affiliation{CEA, INAC-SPINTEC, F-38000 Grenoble}

\begin{abstract}

Using first-principles calculations, we investigated the impact of chromium (Cr) and vanadium (V) impurities on the magnetic anisotropy and spin polarization in Fe/MgO magnetic tunnel junctions. It is demonstrated using layer resolved anisotropy calculation technique, that while the impurity near the interface has a drastic effect in decreasing the perpendicular magnetic anisotropy (PMA), its position within the bulk allows maintaining high surface PMA. Moreover, the effective magnetic anisotropy has a strong tendency to go from in-plane to out-of-plane character as a function of Cr and V concentration favoring out-of-plane magnetization direction for $\sim$ 1.5~nm thick Fe layers at impurity concentrations above 20~\%. At the same time, spin polarization is not affected and even enhanced in most situations favoring an increase of tunnel magnetoresistance (TMR) values.

\end{abstract}
\maketitle
Spin transfer torque Magnetic Random Access memories (STTRAM) are of great interest to microelectronics industry because of their unique combination of assets: 
non-volatility, speed, low power consumption, density, infinite endurance. In particular, STTRAM based on out-of-plane magnetized tunnel junctions (pMTJ) are 
focusing most of the attention because they offer the best efficiency in terms of ratio between their thermal stability factor $\Delta$ which determines the memory 
retention and the write current $I_{c0}$. One of the main goal of microelectronics industry is nowadays to try replacing Dynamic-RAM (DRAM) by STTRAM at dimensions 
below 20nm. To achieve this goal, these memories must exhibit large TMR amplitude (above 200\%), low switching current (below 1 MA/cm$^2$) and high thermal stability 
factor (above 80). Since the thermal stability factor scales with the memory cell area, achieving  $\Delta >80$ becomes increasingly difficult as the cell dimension is 
reduced\cite{1,2}.

In recent publication~\cite{3}, B. Dieny and coworkers proposed to optimize the properties of perpendicular STTRAM for sub-20nm memory applications such as DRAM replacement 
by using double MTJ stacks with anti-parallel polarizing layers. These dual structures have threefold advantages: (i) the storage layer perpendicular magnetic anisotropy
(PMA) is enhanced thanks to the two MgO/CoFe interfaces; (ii) spin transfer torque (STT) efficiency is more than doubled due to the addition of STT contributions originating 
from the two out-of-plane polarizing layers\cite{4};(iii) the stray field from the two polarizing layers on the storage one is minimized so that the two memory states corresponding
to up and down magnetization orientation of the storage layer have same stability. Furthermore, the critical voltages for switching to up or down states are thus 
made more symmetric. To further optimize the downsize scalability of these devices, the effective PMA of the storage layer has to be maximized so that the thermal stability 
factor of the storage layer defined as the ratio between its effective anisotropy volume product and temperature ($\Delta = K_{eff} V/K_BT$ ) remains above 80 which is required to achieve a 
10-years retention in Gbit applications.

The interfacial PMA at magnetic metal/oxide interface \cite{5,6,7,8,9} provides a very convenient way to get simultaneously a large perpendicular anisotropy required for the memory 
retention together with a weak Gilbert damping below 0.015 necessary for low STT switching current. The interfacial PMA at metal/oxide interfaces is remarkably large 
despite the weak spin-orbit of the involved materials. For instance, in Co/AlOx\cite{8}, an interfacial PMA of about 1.45 erg/cm2 for Co/AlOx was measured, which is as 
large as at Pt/Co interface\cite{9}. A PMA of the same order of magnitude was also found at (Co)Fe/MgO interfaces\cite{10,11,12}. This large PMA is usually interpreted in terms
of strong hybridization effect between (Co) Fe-3d orbitals and O-2p orbitals\cite{13,14,15}. However, it has been recently demonstrated that the PMA origin is much more 
complex and goes beyond this simple picture\cite{16}. The overall effective perpendicular anisotropy of the storage layer assumed to be cylindrical and single domain is
given by:

$$\Delta E= K_{eff}V= \left[(K_{s1}+K_{s2}) - \frac{1}{2} (N_Z-N_X) M^2_s.t\right]\pi R^2 $$

where $K_{s1}$ and $K_{s2}$ are the interfacial PMA at the two MgO/CoFe interfaces, $N_Z$  and $N_X$ are the out-of-plane and in-plane demagnetizing coefficients 
which depend on thickness $t$ and radius $R$ , and tend towards $4\pi$ for infinite layer (in CGS units). Since $K_s$ is already closed to its maximum value and 
cannot be further increased, an efficient strategy for enhancing the effective PMA of the storage layer is to minimize the storage layer saturation magnetization 
with however not reducing the interfacial polarization next to the tunnel barrier. In order to accomplish this goal, we propose to introduce Cr and V impurities 
in the Fe layer, i.e. replace Fe by FeX alloy (X= Cr, V). Such low-saturation magnetization alloys have been recently proposed in CoFe(Cr)B/MgO \cite{17} and CoFe(Cr,V)B/MgO\cite{18}
MTJs with in-plane magnetization. In these reports, however, a large decrease of tunnel magnetoresistance (TMR) was observed as a function of impurities concentration 
which is harmful for STTRAM applications. It is therefore necessary to make sure that the reduction of the saturation magnetization with V or Cr addition does not affect 
the interfacial spin polarization responsible for high TMR.  For that purpose, it is proposed to use a composite storage layer in which the bulk of the layer and the 
interfaces are separately optimized. In this case, however, two main questions remain to be addressed: (i) how the TMR will be affected by the introduction of Cr or V in 
the bulk of the storage layer? (ii) How this will impact the interfacial and the overall effective anisotropy? To address these questions, we performed ab-initio 
calculations in order to clarify the influence of single FeX monolayer position within the supercell on its PMA and spin polarization properties. We found that while 
FeX near the interface has a drastic effect in decreasing the interfacial PMA, introducing FeX in the bulk can significantly enhance effective PMA. Furthermore, we 
found that the overall PMA increase as a function of impurity concentration is more efficient in case of vanadium compared to chromium impurities. At the same time, 
introduction of FeX in the bulk of the ferromagnetic electrode is found to have minor effect on the interfacial spin polarization therefore not affecting the TMR amplitude 
of the system.  

Our first-principles calculations are based on density functional theory (DFT) as implemented in the Vienna ab-initio simulation package (VASP) \cite{19} within the framework of 
the projector augmented wave (PAW) potentials \cite{20} to describe electron-ion interaction and generalized gradient approximation (GGA) \cite{21} for exchange-correlation interactions. 
The calculations were performed in three steps. First, fully structural relaxation in shape and volume was performed until the forces become smaller than 0.001 eV/\AA{} 
for determining the most stable interfacial geometries. Next, the Kohn-Sham equations were solved with no spin-orbit interaction taken into account to determine the ground 
state charge distribution of the system. Finally, the spin-orbit coupling was included and the total energy of the system was calculated as a function of the magnetization 
orientation. A $13\times 13\times3$ K-point mesh was used in our calculations. A plane wave energy cut-off equal to 500 eV for all calculations was used and is found to 
be sufficient for our system. We use a2  unit cell with a 5 monolayers (ML) of MgO and 11 ML of Fe as shown in Fig. 1 (a). To find the optimal Fe/FeX/Fe/MgO scenario, 
PMA and interfacial spin polarization was calculated as a function of FeX (1ML) position within the Fe layer and then as a function of its thickness. Here, 
spin polarization is defined as a difference between minority and majority states normalized by the total density of states at the Fermi level, i.e. $P=(n^(\downarrow)-n^(\uparrow)
/n^(\downarrow)+n^(\uparrow))$. The effective anisotropy in our calculation is defined as $K_{eff}=K_s/t_{tot} - 2 \pi M^2_s$  with the second term representing the 
demagnetization energy which favours in-plane anisotropy. However, since we are dealing with thin Fe films, we found that the demagnetization energy calculated from $2\pi M^2_s$
is underestimated by 30\% compared to the one evaluated from the magnetostatic dipole-dipole interaction\cite{16,22}. Therefore, in this work, $K_{eff}$ is defined 
as $K_{eff}=K_s/t_{tot} - E_{demag}$   where $E_{demag}$ is the sum of all the magnetostatic dipole-dipole interactions up to infinity.

\begin{figure}[t]
  \includegraphics  [width=0.5\textwidth]{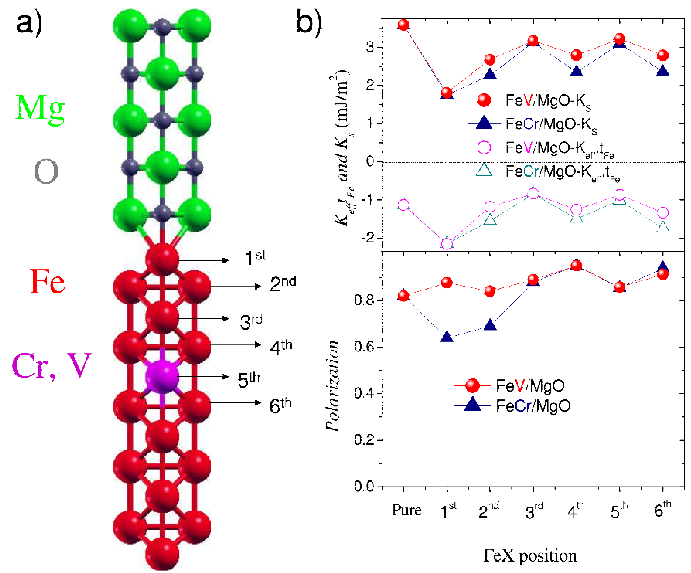}\\
   \caption{(Color online) Schematic of the calculated crystalline structure for a  $\sqrt(2)\times\sqrt{2}$
   unit cell of Fe/FeX/Fe/MgO (X=Cr, V) . Fe,(Cr)V, Mg and O are represented by red, 
   magenta, green and dark balls respectively.(b) (Top panel) Variation of interfacial
   and effective anisotropy as a function of  FeX [1ML] position within the Fe. (b) (Bottom panel) 
   Variation of interface polarization as a function of FeX [1ML] position within the Fe. MgO and 
   total Fe thicknesses are fixed to 5 and 11 monolayers, respectively}\label{fig1}
\end{figure}
\begin{figure*}[t]
  \includegraphics  [width=1.0\textwidth]{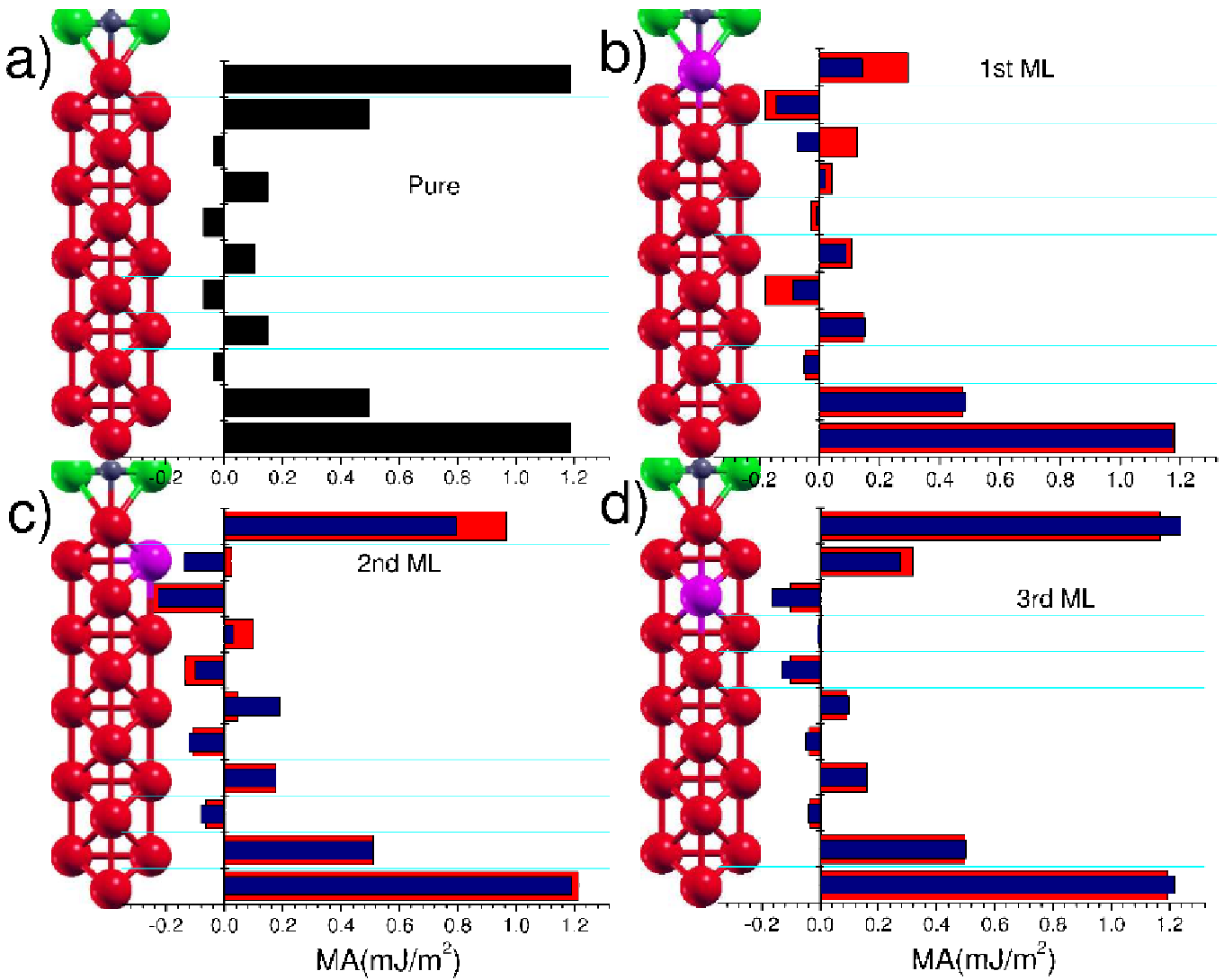}\\
   \caption{(Color online) On-site projected contribution to the magnetic anisotropy for different FeX positions within Fe/FeX/Fe/MgO ( b, c and d) 
   compared to the pure case (a). Red and blue bars represent X=V and X=Cr, respectively}\label{fig2}
\end{figure*}

Figure~\ref{fig1}(b) shows the influence of FeX position on the magnetic anisotropy (top panel) and on the interface polarization (bottom panel). One can see that the FeX 
positioning next to the interface strongly affects the interfacial spin polarization in the case of Cr but much less in the case of V. However, when the FeX is moved 
away from the interface, the interfacial spin polarization is restored indicating that the TMR is not affected when FeX is introduced in the bulk of the layer, i.e. 
farther away from the interface.  Concerning the effective anisotropy, it is found that for the relatively thick Fe/FeX 1ML/Fe layer used here (11 monolayers), the 
demagnetizing energy which favors in-plane anisotropy is always larger than the interfacial PMA so that the effective anisotropy is negative yielding in-plane spontaneous
magnetization. More specifically, introducing the FeX right at the interface (mainly the first two ML) has a drastic effect in decreasing the interfacial PMA values for 
both Cr and V cases. However, introducing the FeX monolayer to the odd ML position in the bulk decreases the in-plane effective anisotropy and enhances it slightly for 
even ML [see Fig.~\ref{fig1}(b)]. The observation of such oscillatory behavior in interfacial PMA has been recently reported experimentally in different systems \cite{23, 24} and is 
attributed to quantum well oscillations in a minority spin d-band at the Fermi level. It has been shown using orbital resolved contribution to the magnetic anisotropy 
that these 2 atomic layer oscillations originates from $\Delta_5$ ( $d_{xz(yz)}$ ) orbitals behaviour \cite{16}.

\begin{figure*}[t]
  \includegraphics  [width=1.0\textwidth]{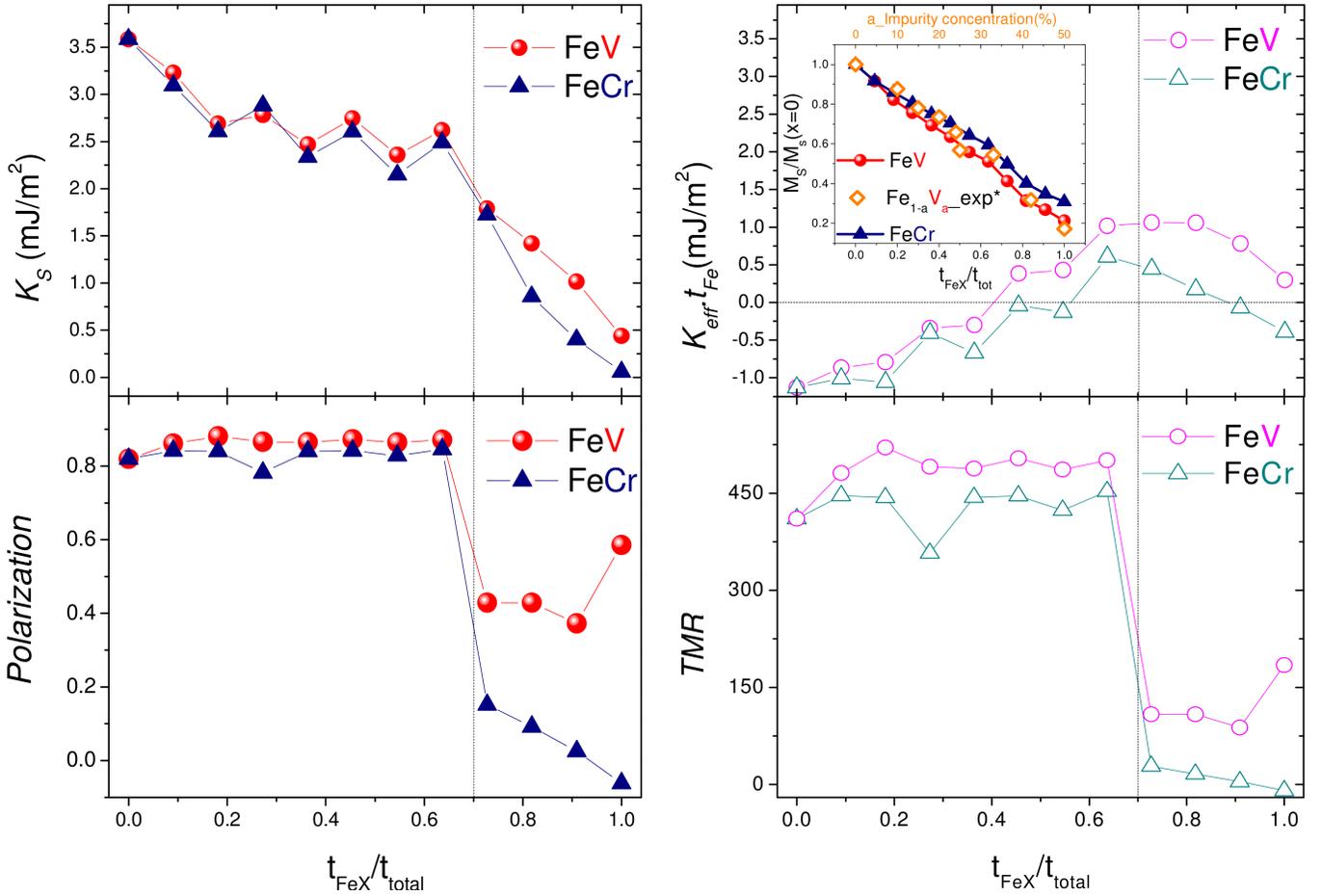}\\
   \caption{(Color online) (Top panel) Dependencies of interfacial (left) and effective (right) anisotropy on 
   relative FeX thickness within the bulk of Fe for Fe/FeX/Fe/MgO (X=Cr, V). Inset show the dependence of the 
   calculated (closed points) magnetic saturation on FeX thickness and the experimental (open points) magnetic
   saturation on V concentration in Fe extracted from reference\cite{26}. t$_{FeX}$/t$_{total}$ here corresponds exactly to 
   V concentration in Fe. (Bottom panel) Dependencies of interface polarization and TMR amplitude on FeX thickness 
   within the bulk of Fe.}\label{fig3}
\end{figure*}

In order to further elucidate the long range influence of V and Cr impurities in the bulk layers on the interfacial PMA, Fig. 2 illustrates the layer resolved 
contributions to the interfacial PMA of the various atomic layers as a function of impurity position in the Fe layer. One can see that in the absence of impurity,
the main contribution to the PMA is localized at the first two layers away from the interface [Fig.~\ref{fig2}(a)]\cite{16} When the impurity is introduced at the interface, 
the interfacial PMA is strongly affected with strong reduction of 1st ML contribution and complete lost of PMA from the 2nd ML [Fig.~\ref{fig2}(b)]. However, moving the 
impurity away from the interface restores the PMA contribution from the interfacial layers as seen in Fig.~\ref{fig2}(c) and (d). Thus, introducing Cr or V in the first 
two layers is harmful for the PMA. In addition, this analysis also allows understanding aforementioned behaviour of PMA shown in Fig.~\ref{fig1}(b). Namely, since 
contributions of the first, second, fourth and sixth MLs are positive [Fig.~\ref{fig2}(a)] i.e. favoring out-of-plane magnetization, introducing the impurity in 
these layers decreases their contribution into the total PMA value [cf. Figs.~\ref{fig1}(b) and~\ref{fig2}(b,c)]. However, the contribution of the third and fifth MLs is 
negative favouring in-plane magnetization. Therefore introducing FeX layer in these odd layers [Fig.~\ref{fig2} (d)] seem not to alter much the value of PMA 
compared to the even case.

We next investigate the influence of increasing the X concentration in the bulk of the Fe layer by varying the relative FeX thickness within the FM layer, 
keeping in each monoatomic FeX plane one X atom for one Fe atom. Figure~\ref{fig3} (top panel) shows the interfacial and effective anisotropy versus FeX thickness. 
The inset of Figure~\ref{fig3} (top panel) shows the calculated magnetic saturation dependence on relative FeX thickness defined as the ratio of the latter to the 
total FM layer thickness, i.e. t$_{FeX}$/t$_{total}$. The inset also show the experimental magnetic saturation as function of V concentration in Fe$_{(1-a)}$V$_a$/MgO
extracted from reference~\cite{25}. As expected for bulk FeCr or FeV alloys, a strong decrease of the magnetization is observed as a function of Cr or V contents, 
actually a decrease by 80\% (respectively 70\%)  at 0K when replacing all the Fe layers by FeX ones. This is known to be due to the formation of a resonant 
virtual bond state in the minority band at the Fermi energy~\cite{26}. Despite the fact that in our calculation we consider a composite material of FeV alloy and Fe, 
unlike in the experiment where they have no control on the V position, our results is in a very good agreement with the experiment. For FeX thickness lower 
than 9 ML (t$_{FeX}$/t$_{total}$ = 0.8), the interfacial anisotropy is decreased by the Cr(V) addition in the bulk of the layer but thanks to the large decrease 
of the magnetization, the demagnetizing energy [$1/2(N_Z-N_X) M^2_s.t_{tot}$] is even more reduced. As a result the effective anisotropy increases. For the 
considered total thickness of 11 magnetic monolayers, a change of sign of the effective anisotropy is even observed (for t$_{FeX}$/t$_{total}$ between 0.5 and 0.8) 
meaning that the effective anisotropy gets reoriented out-of-plane thanks to the FeX bulk substitution. For instance, by comparing the pure case with that of 
t$_{FeX}$/t$_{total}$ = 0.6 of Cr or V, the effective anisotropy is enhanced by more than 1 mJ/m$^2$. However, for thickness above 8 ML, the interfacial anisotropy 
drops drastically from 1.7 mJ/m$^2$ to about zero when the total storage layer is an alloy. Such a strong decrease is due to the fact that the main contribution 
to the PMA comes from the first two ML of the FM. Thus, monotonic increase of the effective PMA stops at t$_{FeX}$/t$_{total}$ = 0.8 and starts decreasing afterwards.
In all cases, V seems to be more efficient than Cr in enhancing the effective perpendicular anisotropy. Not only it gives larger effective anisotropy, but the 
damping is known to be reduced in FeV alloys~\cite{25,27}. V may also help expanding the lattice parameter of Fe thus achieving a better lattice matching with MgO~\cite{28}. 
This may result in a lower density of dislocations which can have a positive impact on the TMR as well as on the resistance to electrical breakdown~\cite{28}.

Finally, we also explore the impact of V and Cr impurities on the TMR amplitude. Indeed keeping in mind the STTRAM application, it is important to make sure that 
the improved retention associated with V substitution in the bulk of the Fe layer does not yield a significant drop of TMR amplitude. In Figure~\ref{fig3} (bottom panel), 
the interfacial spin polarization is plotted. The TMR amplitude was then estimated using Julliere model~\cite{30} as a function of the relative FeX thickness which 
reflects the impurity concentration. A slight enhancement of TMR amplitude is observed when t$_{FeX}$/t$_{total}$ < 0.8, mainly in the case of V impurity which is in 
a good agreement with experimental transport and Andreev reflection measurements showing similar behavior when a small concentration of V were introduced in the bottom 
electrode~\cite{31,32}. For t$_{FeX}$/t$_{total}$ > 0.8, the interfacial spin-polarization and accordingly the TMR amplitude drops quickly both for Cr and V and even reaches zero 
in the Cr case. We can conclude therefore that the proposed introduction of Cr and V impurities into the bulk part of the Fe electrode can significantly enhance the effective 
PMA without destroying the high TMR in Fe/MgO. This is true as long as the FeX is kept away from the interface. From the obtained interfacial anisotropy and $M_s$ values, 
one can show that in double barrier MgO system with slightly thinner V-doped storage layer, a thermal stability factor of 80 can be achieved in these structures down 
to pillar diameter of 10nm. From an electrical point of view, the current required to write can also be calculated. For a thermal stability factor of 80, it is equal 
to 13A in the proposed structure which can be delivered by a transistor (FinFET) 10nm wide. Therefore the MTJ diameter and transistor dimensions are expected to be 
comparable which should allow to achieve a cell size of the order of 6F$^2$ with F=10nm. The figure of merit $\Delta$/Ic of this optimized structure could therefore 
reach 6 $K_BT/\mu A$.

In conclusion, using first-principles calculations, we investigated the impact of Cr and V alloying on the PMA and spin-polarization in Fe/MgO based MTJs. 
Calculations show an increase of effective anisotropy with no reduction of TMR value by introduction of preferably FeV in the bulk of the storage layer while 
keeping pure Fe next to the MgO interface. From these results, scalability down to 10nm should be possible from a magnetic and electrical point of view. However,
processing issues still remain challenging in particular with respect to the dot to dot variability.

\end{document}